\documentstyle[12pt]{article}
\begin{document}
\begin{center}
{\large \bf Nucleon Spin Distributions From} \\
{\large \bf Recent SMC, SLAC and DESY Data} \\
\vspace{5mm}
\underline{Gordon P. Ramsey}$^{1,2}$ and Mehrdad Goshtasbpour$^{3,4}$ \\
\vspace{5mm}
{\small\it
(1) Loyola University of Chicago, Chicago, IL 60626, USA \\
(2) Argonne National Laboratory, IL 60439, USA \\
(3) Shahid Beheshti University, Tehran, Iran \\
(4) Institute for Physics and Mathematics, Tehran, Iran \\}
\end{center}

\begin{center}
ABSTRACT

\vspace{5mm}
\begin{minipage}{130 mm}
\small

We have extracted polarized nucleon distributions from recent data at CERN,
SLAC and DESY. The flavor-dependent valence and sea quark spin distributions are
determined for each experiment. The up and down distributions are comparable,
but the strange sea contribution determined from different experiments do not
agree, even including higher order corrections. Only experiments sensitive to
the polarized gluon and sea will reconcile these differences.

\end{minipage}
\end{center}

Recently, the Spin Muon Collaboration (SMC) group from CERN$^{1}$ and the
experimental groups from SLAC$^{2}$ measured $A_1$ and $g_1$ to low $x$
in deep-inelastic lepton-hadron scattering (DIS) For the proton, neutron and
deuteron. They have improved statistics and minimized the systematic errors.
The measurement of $g_1^{p,n,d}$ provides a means by which we can extract
the polarized quark contributions to proton spin. We have done a detailed
flavor dependent analysis including the QCD corrections and the gluon anomaly.
We use data from each experiment and the sum rules to extract the spin
information.

The polarized valence quark distributions are obtained from the unpolarized
ones by starting with a modified 3-quark model.$^{3}$ The free parameter is
adjusted to satisfy the Bjorken sum rule (BSR).$^{4}$ Using our values
$\langle \Delta u_v\rangle=1.00\pm 0.01$ and
$\langle \Delta d_v\rangle=-.26\pm 0.01$, both the BSR and magnetic moment
ratio $\mu_p/\mu_n$ are satisfied. Thus, the spin contribution from valence
quarks equals $0.74\pm 0.02$. Errors arise from $g_A/g_V$ data and differences
in the unpolarized distributions.

The SU(6) symmetry of the sea is broken by assuming that the polarization of
the heavier strange quarks is suppressed. The sea distributions are then
related by:
\begin{equation}
\Delta \bar{u}(x)=\Delta u(x)=\Delta \bar{d}(x)=\Delta d(x)=[1+\epsilon]
\Delta \bar{s}(x)=[1+\epsilon]\Delta s(x), \label{1}
\end{equation}
where $\epsilon$ is a measure of the increased difficulty in polarizing
the strange quarks.
 
The integrated polarized structure function, $I^{p(n)}\equiv
\int_0^1 g_1^{p(n)}(x)\>dx$, is related to the polarized quark distributions by
\begin{equation}
I^{p(n)}={1\over {18}}(1-\alpha_s^{corr})\langle\bigl[4(1)\Delta u_{tot}+1(4)
\Delta d_{tot}+(\Delta s_{tot})\bigr]\rangle.\label{2}
\end{equation}
where $\alpha_s^{corr}$, has been caluclated to $O(\alpha_s^4)$$^{5}$.
The higher twist corrections$^{6}$ are negligible at the $Q^2$ values of the
data. Additional constraints are provided by the axial-vector current
operators, $A_3$, $A_8$ and $A_0$.

The BSR is a fundamental test of QCD. In terms of the polarized distributions
and equation (1), the BSR can be reduced to:
\begin{equation}
I^p-I^n={1\over 6}\int_0^1 [\Delta u_v(x,Q^2)-\Delta d_v(x,Q^2)]\>dx=
{{A_3}\over 6}(1-\alpha_s^{corr}), \label{3}
\end{equation}                                         
with $I^d=\int_0^1 g_1^d(x)\>dx={1\over 2}[I^p+I^n](1-{3\over 2}\omega_D)$,
where $\omega_D$ is the probability that the deuteron will be in a D-state.
The BSR is used to extract an effective $I^p$ from all data using equation (3).
Then $A_8$, determined by hyperon decay, is given by:
\begin{equation}
A_8=\langle\bigl[\Delta u_v+\Delta d_v+\Delta u_s+\Delta \bar{u}+\Delta d_s+
\Delta\bar{d}-2\Delta s-2\Delta \bar{s}\bigr]\rangle \approx 0.58\pm 0.02.
\label{4}
\end{equation}                                         
$A_0$ is related to the total spin carried by the quarks in the proton.
It can be written in terms of the non-zero axial currents as:
\begin{equation}
A_0= 9(1-\alpha_s^{corr})^{-1}\int_0^1 g_1^p(x)\>dx-{1\over 4}A_8-
{3\over 4} A_3=\langle \Delta q_{tot}\rangle-\Gamma. \label{5}
\end{equation}

The model of $\Delta G$ that is used has a determines the effective quark
distributions through the gluon axial anomaly,$^{7}$ which has
the general form: $\Gamma (Q^2)={{\alpha_s(Q^2)}\over {2\pi}}\int_0^1
\Delta G(x,Q^2)\>dx.$ Each quark flavor is modified by $\Gamma$.
We have used two models for $\Delta G$: (1) $\Delta G=xG$ and (2) $\Delta G=0$.
We believe that present data imply that $\Delta G$ is limited at low $Q^2$.

The orbital angular momentum of the constituents, $L_z$, can be found from:
$J_z={1\over 2}={1\over 2}\langle\Delta q_v\rangle+{1\over 2}\langle
\Delta S\rangle+\langle\Delta G\rangle+L_z.$ Although this does not provide a
constraint on either $\Delta q_{tot}$ or $\Delta G$, it gives an estimate of
the angular momentum component to proton spin.

Equations (1) through (5) are used to extract the flavor dependent information
on the contributions to the proton spin. Results are given in Table I. The E154
and HERMES data are preliminary, as reported in this symposium.

\begin{center}
{\bf Table I: Integrated Polarized Distributions: \\
$\Delta G=xG$ (above line), $\Delta G=0$ (below line)}
\end{center}
$$\begin{array}{cccccc}
  Quantity    &SMC(I^p)  &SMC(I^d)  &E154(I^n)  &E143(I^d) &HERMES \cr
              &          &          &           &          &(I^n)  \cr
  <\Delta u>_{tot} &0.85   &0.82    &0.87       &0.87     &0.90  \cr
  <\Delta d>_{tot} &-.42   &-.43    &-.39       &-.40     &-.36  \cr
  <\Delta s>_{tot} &-.07   &-.10    &-.04       &-.06     &-.02  \cr
     I^p        &0.136   &0.129     &0.134      &0.131    &0.135 \cr
  <\Delta q>_{tot} &0.36   &0.29    &0.45       &0.41     &0.52 \cr
     L_z        &-.14    &-.11      &-.18       &-.15     &-.22 \cr
 ------& -----& -----& ------& -----& ----- \cr
 <\Delta u>_{tot} & .83     & .80    & .85    & .84   & .88   \cr
 <\Delta d>_{tot} & -.44    & -.45   & -.41   & -.43  & -.39  \cr
 <\Delta s>_{tot} & -.09    & -.12   & -.07   & -.08  & -.04  \cr
 <\Delta q>_{tot} & 0.30    & 0.23   & 0.37   & 0.33  & 0.45 \cr
     L_z          & 0.35    & 0.39   & 0.32   & 0.35  & 0.28
\end{array}$$ 

From the results in Table I, we can draw the following conclusions:

(1) The naive quark model is not sufficient to explain the proton's spin
characteristics, since the total quark contribution to proton spin falls
between about $1\over 4$ and $1\over 2$. The uncertainties due to data and
the choice of $\Delta G$ are comparable.

(2) The up and down total contributions to proton spin all agree to within a
few percent. Most data, including that of the proton and deuteron, imply a
larger polarized sea with the strange sea polarized greater than the positivity
bound.$^{8}$ The strange sea contribution is the most uncertain of all the
flavors. Higher twist corrections do not reconcile these differences.

(3) This analysis is consistent with a small anomaly correction. Specifically,
a larger anomaly term from a greater $\Delta G$ implies that the strange sea
would be positively polarized, while the other flavors are negatively
polarized. Since there is no obvious mechanism that allows selective
polarization of different flavors, we conclude that these data imply that
$\Delta G$ is of small to moderate size.

(4) The orbital angular momentum extracted from data is rather small, but
could be positive or negative, depending on the size of the polarized
gluon distribution.

(5) The extracted value for $I^p$ is comparable for all the data and well
within experimental uncertainties. This indicates the validity of the BSR.
 
These experiments along with theoretical progress in calculating higher order
QCD corrections, have allowed us to narrow the range of the spin contributions
for each flavor. They have probed to smaller $x$ values, while decreasing the
statistical and systematic errors. The main differences are the strange sea
spin content and the size of $\Delta G$. There are a number of technologically
feasible experiments that would supply more information about these
distributions. Detailed summaries can be found in reference 9.

{\small\begin{description} 
\item{[1]} 
B. Adeva, {\it et.al.}, Phys. Lett. {\bf B302} (1993) 533 and Phys. Lett.
{\bf B320} (1994) 400; D. Adams, {\it et.al.}, Phys. Lett. {\bf B329} (1994)
399. 
\item{[2]} 
P.L. Anthony, {\it et.al.}, Phys. Rev. Lett. {\bf 71} (1993) 959; K. Abe,
{\it et.al.}, Phys. Rev. Lett. {\bf 74} (1995) 346. 
\item{[3]} 
J.-W. Qiu, {\it et. al.}, Phys. Rev. {\bf D41} (1990) 65 and M. Goshtasbpour
and G. Ramsey, hep-ph 9512250.
\item{[4]} 
J.D. Bjorken, Phys. Rev. {\bf 148} (1966) 1467. 
\item{[5]} 
S.A. Larin, Phys. Lett. {\bf B334} (1994) 192
\item{[6]} 
E. Stein, {\it et. al.}, Phys. Lett. {\bf B353} (1995) 107.
\item{[7]} 
A.V. Efremov and O.V. Teryaev, JINR Report E2-88-287 (1988); R.D. Carlitz,
J.C. Collins, and A.H. Mueller,  Phys. Lett. {\bf B214} (1988) 229.
\item{[8]} 
G. Preparata, P.G. Ratcliffe and J. Soffer, Phys. Lett. {\bf B273} (1991)
306.
\item{[9]} 
G.P. Ramsey, Particle World, {\bf 4}, No. 3, (1995).
\end{description}}
\end{document}